\documentclass{svmult}

\usepackage{makeidx}     

\usepackage{graphicx}    

\usepackage{multicol}    

\makeindex             

\begin{document}

\title*{Ideal Gas-Like Distributions in Economics: Effects
of Saving Propensity}

\author{Bikas K. Chakrabarti and Arnab Chatterjee}

\institute{Saha Institute of Nuclear Physics, 1/AF Bidhan Nagar, Kolkata-700064, India
\texttt{bikas@cmp.saha.ernet.in}}

\maketitle

We consider the ideal-gas models of trading markets, where each agent is 
identified with a gas molecule and each trading as an elastic or 
money-conserving (two-body) collision. Unlike in the ideal gas, we introduce
saving propensity $\lambda$ of agents, such that each agent saves a fraction
$\lambda$ of its money and trades with the rest. We show the steady-state 
money or wealth distribution in a market is Gibbs-like for $\lambda=0$, has
got a non-vanishing most-probable value for $\lambda \ne 0$ and Pareto-like
when $\lambda$ is widely distributed among the agents. We compare these results
with observations on wealth distributions of various countries.

\section{Introduction}
\label{sec:1}
I should start by thanking the organisers of this Symposium
for inviting me (BKC) and also perhaps more importantly for highlighting our
humble efforts in Kolkata (erstwhile Calcutta), by mentioning in their
(symposium) homepage that ``The term econophysics was ... first used
in 1995 at an international conference ... in Calcutta'' \cite{niki}.
Indeed Kolkata is the formal birth place of this new term. It was
first used by Stanley \cite{stanley} in the second of the Statphys-Kolkata
series of conferences (being held over more than a decade now 
\cite{statphys}), where several important developments in this 
interdisciplinary emerging field had been reported by many physicists.

In an earlier conference in Kolkata in 1994, many leading
Indian economists from the Indian Statistical Institute and physicists
met and discussed about the possible formulations of some economic
problems and their solutions using tricks from physics \cite{kol95}.
In one of these papers \cite{bkcmarj}, possibly the first 
published joint paper with both physicist and economist Indian coauthors,
the possibility of a kinetic theory of (ideal) gas-like
model of trading in the market was discussed. Among other things, it
tried to identify, from the known effects of various fiscal policies,
equivalence of the (kinetic) energy of the gas molecules (money) and
the temperature (average money in the market). Such a ``finite temperature''
market model and the corresponding distributions were also
noted by others \cite{olids, dy1}. With the possibility of putting 
more than one agent in the same (micro) state, identified by the price or money
income of the agent in the market, the likely distribution was concluded
there to be Bose-Einstein like, rather than Gibbs like \cite{bkcmarj}.
These studies of course had the limitation of absence of
any comparison with real income distributions (in any market or country).
In a recent paper by Dragulescu and Yakovenko \cite{dy1} a simple (trading)
market model was developed with fixed (total) money and number of agents
in the market. Random two-agent exchanges (with local money conservation)
lead to Gibbs-like steady income distribution. This was also confirmed
by simple numerical simulations. Modifications due to savings was
studied simultaneously \cite{acbkcepjb}. In a very
recent review \cite{hayes}, a popular introduction to these developments
is given. 

Saving propensity among the agents, a very selfish and local 
feature of the tradings, introduce in effect some global co-operative
feature (cf. \cite{acbkcepjb}). We show that a fixed and uniform saving
propensity of all the agents in the market shifts the most-probable money of 
the distribution away from zero (as given by Gibbs for zero savings), while
a random distribution of saving propensity among the population can give the
Pareto (power) law \cite{pareto}

\begin{equation}
\label{paret}
P(m) \sim m^{-(1+\nu)} 
\end{equation}

\noindent for the wealth or money ($m$) distribution.
We intend to discuss here in brief the effects of various kinds of savings on
the ideal gas-like money distributions in the above-mentioned market
models, and compare our observations with those from real markets.

\section{An Ideal Gas-Like Market Model}
\label{sec:2}

Let us consider a simple model of a closed economic system
where the total amount of money $M$ and the total number $N$ of agents are
fixed. No development (production) or migration (death/birth of agents)
occurs and the only economic activity is confined to trading. Each agent
$i$, individual or a corporate, possess a money $m_i(t)$ at (discretised)
time $t$. Time changes after each trading. In any trading, two randomly
chosen agents $i$ and $j$ exchange their money such that their total
money is (locally) conserved and none ends up with negative money (debt not
allowed):

\begin{equation}
\label{consv}
m_i(t) + m_j(t) = m_i(t+1) + m_j(t+1)
\end{equation}

\noindent where $m_i(t) \ge 0$ for all $i$ and $t$; $\sum_{i=1}^N m_i = M$.
Since money is conserved, in the steady state ($ t \rightarrow \infty$), the
probability $P(m)$ of the density of people with money $m$ will satisfy

\begin{equation}
\label{prob}
P(m_1)P(m_2) = P(m_1 + m_2)
\end{equation}

\noindent which corresponds to the Gibbs distribution \cite{dy1, acbkcepjb}

\begin{equation}
\label{gibbs}
P(m) = \left( 1/T \right) \exp \left( -m/T \right); ~~~ T=M/N.
\end{equation}

Numerical simulations in the model also confirmed the steady state distribution
of money, no matter what initial distribution of money the agents had, to
be Gibbs' one: after sufficient number of tradings, most of the agents end
up with very little money! This result is quite robust (and realistic too!).
In fact, several variations of the trading, and of the `lattice' (on which
the agents can be put and each agent trade with its `lattice neighbours' only),
whether compact, fractal or small-world \cite{olids}, leaves the distribution 
unchanged.
Some other variations like random sharing of an amount $2m_2$ only (not of
$m_1 + m_2$) when $m_1 > m_2$ (trading at the level of lower economic
class), lead to even drastic situation: all the money in the market drifts
to one agent and the rest of the agents all become truely pauper 
\cite{hayes,chaxijmpc}! Attempts have also been made \cite{tsalreiss} to get
Pareto-like power-law distribution here with changed definition of entropy
or the conservation law (cf. eqn. (\ref{prob})).

\vskip 0.3cm
\begin{figure}
\centering
\includegraphics[height=5cm,width=8cm]{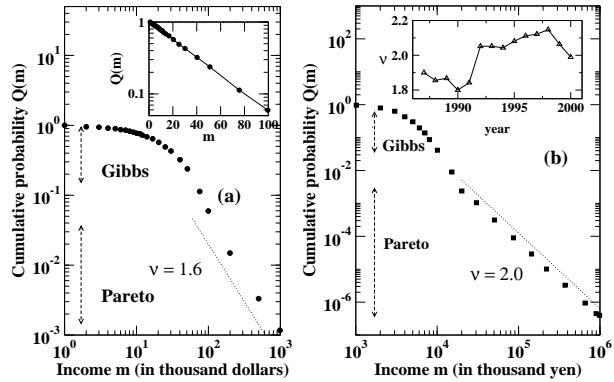}
\caption{Cumulative distribution $Q(m)=\int_m^{\infty}P(m)dm$ of individual 
income (a) in USA in 1997 (data taken from \cite{dy2dy3}), 
(b) in Japan in 2000 \cite{fuji}. The inset in (a) shows the validity of Gibbs
law for low income and in (b) the recently observed variations of the 
Pareto index $\nu$ (for higher income group) in Japan. Typically, 10\% of the
higher income group people possess about 40\% of the wealth in any economy
and follow Pareto law (\ref{paret}) with index $\nu$ in the range $1$ to $2$
\cite{olids}.}
\label{fig:1}
\end{figure}

Chakrabarti and Marjit \cite{bkcmarj} argued for the Bose-Einstein like 
distribution (rather than Gibbs) in such a market (with the temperature 
$T$ similarly identified with the average money $M/N$ per agent), as one can 
put more than one agent in the same economic state (specified by the income) 
and the maximisation of the consequent entropy. For the possibility of adding
and subtracting agents into/from the market, one similarly needs (negative)
``chemical potential'' which becomes zero at a finite temperature or money
level in the market, when the ``Bose condensed'' fraction of the agents
will fall out of the market distribution and might be identified as
unemployed.

The real income distributions did not indicate so far anything like
the Bose distribution; rather considerable evidences support the
possibility of Gibbs like distribution (\ref{gibbs}) in the income 
(almost for 90\% of the low-income range) of various countries 
(see e.g., \cite{dy2dy3}, see also data in \cite{fuji}; Fig. 1). 

\section{Model with fixed saving propensity of the agents}
\label{sec:3}

Here we assume \cite{acbkcepjb} that each economic agent \( i \)
saves a fraction \( \lambda  \) of its money \( m_{i}(t) \) before
the trading at time \( t \). We again assume that an agent's money
is non-negative and no debt is permitted. Let us now consider an arbitrary
pair of agents \( i \) and \( j \), who get engaged in a trade, and
their money \( m_{i}(t) \) and \( m_{j}(t) \) before the trade
change respectively to 
\begin{equation}
\label{delm}
m_{i}(t+1)=m_{i}(t)+\Delta m;\quad m_{j}(t+1)=m_{j}(t)-\Delta m 
\end{equation}

\begin{equation}
\label{eps}
{\rm where}\;\;\;\;\; \Delta m=(1-\lambda )[\epsilon \{m_{i}(t)+m_{j}(t)\}-m_{i}(t)];
\end{equation}

\noindent with \( \epsilon  \) as any random fraction. As may be checked by
straight-forward substitution, this kind of trading again satisfies
eqn. (2), while each agent saves a fixed fraction \( \lambda  \)
of its money before the trade and exchanges randomly (with fraction
\( \epsilon  \)) the rest of the money.

One finds here that at \( \lambda =0 \) the market becomes non-interacting
and the steady state money distribution becomes the Gibbs' one.
For any nonvanishing \( \lambda  \), the equilibrium distribution
becomes asymmetric Gibbs-like (see Fig. 2a) with the most-probable
money \( m_{p} \) per agent (corresponding to the peak in \( P(m) \))
shifting away from \( m=0 \) (for \( \lambda =0 \)) to \( M/N \)
as \( \lambda \rightarrow 1 \) \cite{acbkcepjb}. This self-organising
feature of the market, induced by sheer self-interest of saving by each agent
without any global perspective, is very significant as the fraction of paupers 
decreases with saving fraction $\lambda$ and most people end up with the 
average money in the 
market (socialists' dream achieved with just people's self-interest of saving)! 
Interestingly, self-organisation also occur in such market models when there
is restriction in the commodity market \cite{acspbkc}.
Although this fixed saving propensity does not give yet the Pareto-like 
power-law distribution, the Markovian nature of the scattering or trading
processes (eqn. (\ref{prob})) is lost and the system becomes co-operative. 
Indirectly through \(\lambda\), the agents get to know (start interacting with)
each other and the system co-operatively self-organise towards a most probable
distribution (\(m_p \ne 0\)).

\section{Model with random saving propensity of the agents}
\label{sec:4}

We now consider a market again with fixed \( N \) and \( M \) but
with random saving propensity \( \lambda _{i} \) (\( 0\leq \lambda _{i}< 1 \))
fixed or ``quenched'' for each agent (\( \lambda _{i} \)
are independent of trading or \( t \), but vary randomly from agent to agent)
\cite{acnccm}. One again follows the same trading rules as mentioned
in the previous section (eqn. (\ref{delm})), except that

\begin{equation}
\Delta m=\epsilon(1-\lambda _{j})m_{j}(t)-(1-\lambda _{i})(1 - \epsilon)m_{i}(t)
\end{equation}

\noindent here; \( \lambda _{i} \) and \( \lambda _{j} \) are the saving 
propensities of agents \( i \) and \( j \).
We first take a market with \( N \) agents, each having a fixed
saving propensity \( \lambda  \)
distributed  independently, randomly and uniformly (white) within an interval
\( 0 \) to \( 1 \).  Having assigned each agent \( i \) the saving
propensities \( \lambda _{i} \), and starting with an arbitrary initial
(uniform or random) distribution of
money among the agents, we start the tradings. At each time, two agents
are randomly selected and the money exchange among them occurs, following
the above mentioned scheme. We check for the steady state, by looking at the
stability of the money distribution \( P(m) \) in successive
Monte Carlo steps \(t\).

\vskip 0.3cm
\begin{figure}
\centering
\includegraphics[height=5cm,width=5cm]{fig2a.eps}\hskip 0.3cm \includegraphics[height=5cm,width=6cm]{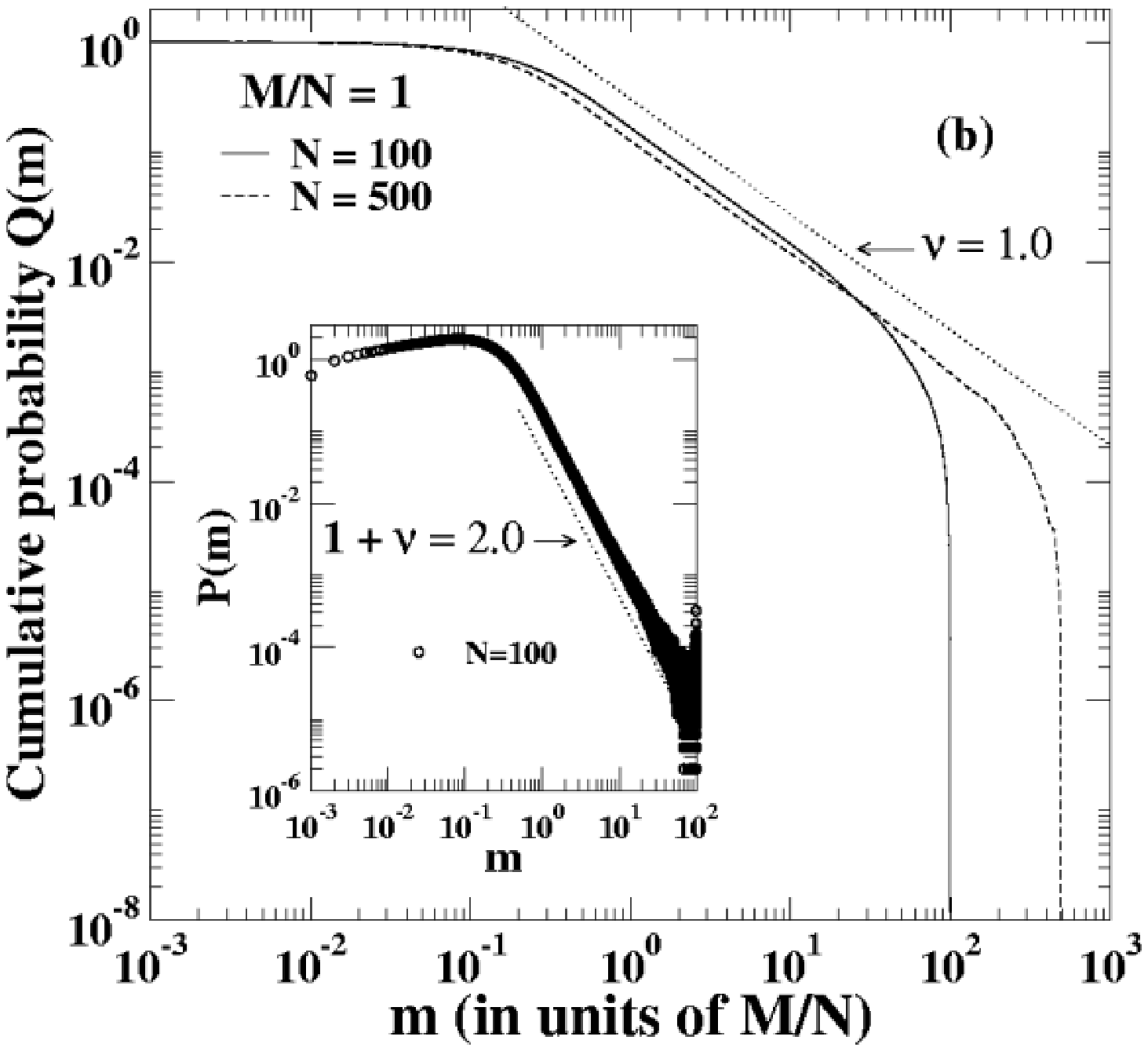}
\caption{Simulation results in the model with saving propensities: (a) Stable 
distribution of money 
$P(m)$ vs. $m$ for different (but fixed for all agents) saving propensity 
$\lambda$ of agents ($N=100$). The inset shows that the 
most-probable money $m_p$ of any agent in the market shifts with $\lambda$ 
($m_p =0$ for $\lambda = 0$; Gibbs law). (b) Stable
cumulative probability distribution of money ($Q(m)=\int_m^{\infty} P(m)dm$) in a 
model market with quenched distribution of $\lambda$ within the range 
$0 \le \lambda < 1$; $N=100, 500$. The dotted line corresponds to a 
power law decay with $\nu \simeq 1.0$. The inset shows the stable probability 
distribution of money $P(m)$ vs. $m$, averaged over $10^6$ initial 
configurations.}
\label{fig:2}
\end{figure}

In the inset of Fig. 2b, we show the money distribution \( P(m) \)
vs. \( m \) (in units of \( M/N \)) for \( N=100 \), \( M/N=1 \),
after averaging over \( 10^6 \) initial configurations (\( \lambda _{i} \)
distribution among the agents) at \( t/N \) =10,000. There
is an initial growth of \( P(m) \) from \( m=0 \), which quickly
saturates and then a long range of power-law decay in \( P(m) \) for large
\(m\) values (for less than 10\% of the population $N$ in the market) is 
observed (for more than two decades in \(m\)). This decay, when fitted to 
Pareto law (\ref{paret}), gives \( \nu = 1.03 \pm 0.03 \).

\section{Summary and Conclusions}
\label{sum}

We have considered ideal-gas models of trading markets. In these models, we 
introduce
saving propensity $\lambda$ of agents, such that each agent saves a fraction
$\lambda$ of its money and trades with the rest. We show the steady-state 
money or wealth distribution $P(m)$ in the market is that of Gibbs 
(\ref{gibbs}) for $\lambda=0$, has got a non-vanishing most-probable value 
for $\lambda > 0$ (but fixed for all agents), and one gets Pareto  
distribution (\ref{paret}) with $\nu \simeq 1.0$ 
when $\lambda$ is widely distributed among the agents. These results in simple 
ideal-gas like market models also compare well with real market observations.

\printindex
\end{document}